\documentclass[12pt]{article}
\usepackage{epsfig}
\newcommand{\be}{\begin{equation}}
\newcommand{\ee}{\end{equation}}
\newcommand{\bea}{\begin{eqnarray}}
\newcommand{\eea}{\end{eqnarray}}
\newcommand{\ba}{\begin{array}}
\newcommand{\ea}{\end{array}}
\newcommand{\beas}{\begin{eqnarray*}}
\newcommand{\eeas}{\end{eqnarray*}}
\newcommand{\bes}{\begin{equation*}}
\newcommand{\ees}{\end{equation*}}

\def\Box{\hbox{$\rlap{$\sqcup$}\sqcap$}}

\def\i2           {\mbox{$\frac{i}{2}$}}

\begin{document}
\title{\bf Quantum Gravitational Corrections to the Real Klein-Gordon Field in the Presence of a Minimal Length}

\author{S. K. Moayedi $^{a}$\thanks{E-mail: s-moayedi@araku.ac.ir}\hspace{1mm}
, M. R. Setare $^{b}$ \thanks{E-mail: rezakord@ipm.ir}\hspace{1mm}
and
 H. Moayeri $^{a}$\thanks{E-mail: h-moayeri@phd.araku.ac.ir}\hspace{1mm} \\
$^a$ {\small {\em  Department of Physics, Arak University,
38156-879
Arak, Iran}}\\
$^{b}${\small {\em Department of Science, Payame Noor University, Bijar, Iran }}\\
}
\date{\small{}}
\maketitle

\begin{abstract}
\noindent

The (D+1)-dimensional $(\beta,\beta')$-two-parameter
Lorentz-covariant deformed algebra introduced by Quesne and
Tkachuk [C. Quesne and V. M. Tkachuk, J. Phys. A: Math. Gen.
\textbf {39}, 10909 (2006).], leads to a nonzero minimal
uncertainty in position (minimal length). The Klein-Gordon
equation in a (3+1)-dimensional space-time described by
Quesne-Tkachuk Lorentz-covariant deformed algebra is studied in
the case where $\beta'=2\beta$ up to first order over deformation
parameter $\beta$. It is shown that the modified Klein-Gordon
equation which contains fourth-order derivative of the wave
function describes two massive particles with different masses.
We have shown that physically acceptable mass states can only
exist for $\beta<\frac{1}{8m^{2}c^{2}}$ which leads to an
isotropic minimal length in the interval
$10^{-17}m<(\bigtriangleup X^{i})_{0}<10^{-15}m$. Finally, we
have shown that the above estimation of minimal length
is in good agreement with the results obtained in previous investigations. \\
\end{abstract}
\section{Introduction}
One of the most important problems in theoretical physics is to
unify general relativity and quantum mechanics. Together they are
able to describe almost all known phenomena from the scale of
subatomic particles all the way up to the rotations of galaxies
and even the large-scale structure of the universe. On the other
hand, according to fundamental principles of quantum mechanics
the position and momentum of a particle cannot be measured
simultaneously. The uncertainties of position and momentum are
related by Heisenberg's uncertainty relation $\Delta x\Delta p
\geq \frac{\hbar}{2}$. An important consequence is that in order
to probe arbitrarily small length-scales, one has to use probes
of sufficiently high energy, and thus momentum. At present, there
is theoretical evidence that in quantum gravity there might exist
a minimal observable distance on the order of the Planck length,
$l_{P}=\sqrt{\frac{G\hbar}{c^{3}}}\simeq 1.6\times 10^{-35}\; m$,
where G is the Newton's constant. From string theory we know that
a string cannot probe distances smaller than its length. The
existence of such a minimal observable length which is motivated
from perturbative string theory, quantum gravity and black hole
gedanken experiments, leads to a generalization of Heisenberg
uncertainty principle. This generalized or gravitational
uncertainty principle (GUP) can be written as
\begin{equation}
\triangle X\triangle P\geq\frac{\hbar}{2}[1+\beta(\triangle
P)^{2}+...],
\end{equation}
where $\beta$ is a positive parameter [1-7]. It is clear that in
Eq. (1), $\Delta X$ is always larger than $(\triangle X)_{min}
=\hbar\sqrt\beta$. Nowadays, physicists are trying to reformulate
the quantum field theory in the presence of a minimal observable
length and there is hoping that this approach causes unwanted
divergencies can be eliminated or modified in quantum field theory [4].\\
Our paper is organized as follows. In Sec. 2, we will obtain a
generalized Klein-Gordon equation of the fourth-order, describing
particles with spin-0 based on the Lorentz-covariant deformed
algebra with minimal length which was introduced by Quesne and
Tkachuk in Refs. [2,3]. In Sec. 3, the solutions of the
generalized wave equation for free motion of a Klein-Gordon
particle have been obtained and it has been shown that these
solutions are associated with two different mass states. Then, we
will propose a direct method for obtaining the minimal length. We
find that the minimal length is of the order of $10^{-16}m$. Our
estimation of the minimal length is in good agreement with the
results obtained in the papers [8-13]. The conclusions are
presented in Sec. 4. Two appendices conclude this paper. Appendix
A presents the Lagrangian formulation of the real Klein-Gordon
field in the presence of a minimal length. In appendix B, the
Feynman propagator in momentum space for the real Klein-Gordon
field in the presence of a minimal length has been obtained and
it has been shown that the generalized real Klein-Gordon field
possesses one physical state with the mass $m_{-}$, and one Weyl
ghost with the mass $m_{+}$. SI units are used throughout this
paper.

\section{ Generalized Klein-Gordon Equation}
In 2006, Quesne and Tkachuk introduced a Lorentz-covariant
deformed algebra which describes a (D+1)-dimensional quantized
space-time [2,3]. This algebra in (3+1)-dimensional space-time is
characterized by the following modified commutation relations
\begin{equation}
[X^{\mu},P^{\nu}]=-i\hbar(g^{\mu\nu}(1-\beta
P_{\rho}P^{\rho})-\beta'P^{\mu}P^{\nu}),
\end{equation}
\begin{equation}
[X^{\mu},X^{\nu}]=i\hbar\frac{2\beta-\beta'-(2\beta+\beta')\beta
P_{\rho}P^{\rho}}{1-\beta
P_{\rho}P^{\rho}}(P^{\mu}X^{\nu}-P^{\nu}X^{\mu}),
\end{equation}
\begin{equation}
[P^{\mu},P^{\nu}]=0,
\end{equation}
where $\mu,\nu =0,1,2,3$ and $\beta,\beta'$ are two deformation
parameters which are assumed positive $(\beta,\beta'\geq 0)$. In
terms of length (L), mass (M), and time (T) the deformation
parameters $\beta$ and $\beta'$ have the same dimensions
$M^{-2}L^{-2}T^{2}$. Also, $X^{\mu}$ and $P^{\mu}$ are deformed
position and momentum operators and
$g_{\mu\nu}=g^{\mu\nu}=diag(1,-1,-1,-1)$. By using Eq. (2) and
Schwarz inequality the uncertainty relation for position and
momentum by assuming that $\triangle P^{i}$ is isotropic
$(\triangle P^{i}=\triangle P,\hspace{2mm} i=1,2,3)$ becomes
 \begin{equation}
\triangle X^{i}\triangle
P\geq\frac{\hbar}{2}\left|1-\beta\left\{\langle(P^{0})^{2}\rangle
-3(\triangle P)^{2}-\sum^{3}_{j=1}\langle
P^{j}\rangle^{2}\right\}+\beta'\left[(\triangle P)^{2}+\langle
P^{i}\rangle^2\right]\right|.
\end{equation}
Hence, we arrive at an isotropic absolutely smallest uncertainty
in position given by
\begin{equation}
(\triangle
X^{i})_{0}=\hbar\sqrt{(3\beta+\beta')\left[1-\beta\langle
(P^{0})^{2}\rangle\right]}\hspace{2mm}, \hspace{5mm}i\in
\{1,2,3\}.
\end{equation}
In this study, we consider the special case $\beta'=2\beta$,
wherein the position operators $X^{\mu}$ commute to first order
in $\beta$. In such a linear approximation, the Lorentz-covariant
deformed algebra reads
\begin{equation}
[X^{\mu},P^{\nu}]=-i\hbar(g^{\mu\nu}(1-\beta
P_{\rho}P^{\rho})-2\beta P^{\mu}P^{\nu}),
\end{equation}
\begin{equation}
[X^{\mu},X^{\nu}]=0,
\end{equation}
\begin{equation}
[P^{\mu},P^{\nu}]=0.
\end{equation}
It is straightforward to verify that the following
representations satisfy the Eqs. (7)-(9), at the first order in
$\beta$,
\begin{equation}
X^{\mu}=x^{\mu},
\end{equation}
\begin{equation}
P^{\mu}=(1-\beta p^{2})p^{\mu},
\end{equation}
where $x^{\mu}, p^{\mu}=i\hbar\frac{\partial}{\partial
x_{\mu}}=i\hbar\partial^{\mu}$ are position and momentum operators
in ordinary relativistic quantum mechanics, and
$p^{2}=p_{\alpha}p^{\alpha}$. The Klein-Gordon equation for a
spin-0 particle with rest mass $m$ is [14]
\begin{equation}
p_{\mu}p^{\mu}\Phi-m^{2}c^{2}\Phi=0.
\end{equation}
In particle physics the real Klein-Gordon field describes
electrically neutral spin-0 particles such as $\pi^{0}-meson$.
Now the Klein-Gordon equation (12) should be written in
generalized form. For this purpose, the momentum operator
$p^{\mu}$ must be replaced by deformed momentum operator
$P^{\mu}$ from Eq. (11), hence we have
\begin{equation}
(1-\beta p^{2})p_{\mu}(1-\beta p^{2})p^{\mu}\Phi-m^{2}c^{2}\Phi=0.
\end{equation}
Neglecting terms of order $\beta^{2}$, the generalized
Klein-Gordon equation (13) takes the form
\begin{equation}
\Box\Phi+2\beta
\hbar^{2}\Box\Box\Phi+(\frac{mc}{\hbar})^{2}\Phi=0,
\end{equation}
where $\Box\equiv\partial_{\mu}\partial^{\mu}$ is the
d'Alembertian operator. The term $2\beta\hbar^{2}\Box\Box\Phi$ in
Eq. (14) can be considered as a quantum gravitational correction.
The wave equation (14) is a fourth-order relativistic wave
equation that in the limit of $\beta\rightarrow 0$ turns in to
the ordinary Klein-Gordon equation. In appendix A we have studied
the Lagrangian formulation of the real Klein-Gordon field in the
presence of a minimal length.

\section{Plane-Wave Solutions of the Generalized Klein-Gordon Equation}
In this section, we will obtain the plane-wave solutions of the
generalized Klein-Gordon equation (14). We try to find a
plane-wave solution of (14):
\begin{equation}
\Phi=A e^{-ik.x},
\end{equation}
where $A\neq 0$ is a complex constant. Equation (15) is a
solution of (14) if
\begin{equation}
2\beta\hbar^{2}(k^{2})^{2}-k^{2}+\left(\frac{mc}{\hbar}\right)^{2}=0,
\end{equation}
where
\begin{equation}
k^{2}=k_{\mu}k^{\mu}=\left(\frac{\omega}{c}\right)^{2}-\vec{k}.\vec{k}\hspace{2mm}.
\end{equation}
By solving Eq. (16) with respect to $k^{2}$, we obtain two
different values for $k^{2}$ as
\begin{equation}
k^{2}_{+}=\left(\frac{m_{+} c}{\hbar}\right)^{2},
\end{equation}
\begin{equation}
k^{2}_{-}=\left(\frac{m_{-} c}{\hbar}\right)^{2},
\end{equation}
where the non-degenerate effective masses $m_{+}$ and $m_{-}$ are
defined as
\begin{equation}
m_{+}=\frac{(1+2\sqrt{2\beta}mc)^{\frac{1}{2}}+(1-2\sqrt{2\beta}mc)^{\frac{1}{2}}}{2\sqrt{2\beta}c},
\end{equation}
\begin{equation}
m_{-}=\frac{(1+2\sqrt{2\beta}mc)^{\frac{1}{2}}-(1-2\sqrt{2\beta}mc)^{\frac{1}{2}}}{2\sqrt{2\beta}c}.
\end{equation}
From the standpoint of quantum mechanics, Eqs. (20) and (21)
indicate that our field is associated with particles having the
effective masses $m_{+}$ and $m_{-}$. To avoid particles of
complex mass, Eqs. (20) and (21) require that
\begin{equation}
\beta<\frac{1}{8m^{2}c^{2}}.
\end{equation}
Note that at $\beta=\frac{1}{8m^{2}c^{2}}$ both effective masses
are equal, i.e., $m_{+}=m_{-}=m\sqrt{2}$. Using Eqs. (17)-(19) we
arrive at the following generalized energy-momentum relations
\begin{equation}
E_{p}^{(+)2}=m^{2}_{+}c^{4}+c^{2}\left|\vec{p}\right|^{2},
\end{equation}
\begin{equation}
E_{p}^{(-)2}=m^{2}_{-}c^{4}+c^{2}\left|\vec{p}\right|^{2},
\end{equation}
where $E_{p}^{(\pm)}=\hbar\omega_{k}^{(\pm)}$. The effective
masses $m_{+}$ and $m_{-}$ in Eqs. (20) and (21) in the first
order over deformation parameter $\beta$ can be written as
\begin{equation}
m_{+}=\frac{1}{\sqrt{2\beta}\; c}-\frac{m^{2}}{2}\sqrt{2\beta}\;
c,
\end{equation}
\begin{equation}
m_{-}=m+\beta m^{3}c^{2}.
\end{equation}
Hence for $\beta\rightarrow 0$ the effective mass $m_{-}$ in Eq.
(26) reduces to the ordinary mass $m$. Inserting (26) into the
equation (24) we find the following generalization for
energy-momentum relation
\begin{equation}
E_{p}^{(-)2}=m^{2}c^{4}+c^{2}\left|\vec{p}\right|^{2}+2\beta
m^{4}c^{6}.
\end{equation}
When $\beta=0$, equation (27) will be converted into the
well-known Einstein energy-momentum relation in the special
relativity. On the other hand, the effective mass $m_{+}$ in Eq.
(25) diverges for small $\beta$. In appendix B, we have shown
that the real Klein-Gordon field in the presence of a minimal
length possesses one physical state with the effective mass
$m_{-}$, and one Weyl ghost with the effective mass $m_{+}$ and
hence the other generalized energy-momentum relation for the
effective mass $m_{+}$ in Eq. (23) is entirely new and for
$\beta=0$, it has no analog in the special theory of relativity.
The general solution of Eq. (14) is a superposition of
plane-waves as
\begin{eqnarray}
\Phi(x)=\sum_{\vec{k}}\left(\frac{\hbar
c^{2}}{2V\omega^{(-)}_{k}}\right)^{\frac{1}{2}}\left[a(\vec{k})\;exp\left(-\frac{i}{\hbar}(E^{(-)}_{p}t-\vec{p}.\vec{r})\right)
+a^{\dag}(\vec{k})\;exp\left(\frac{i}{\hbar}(E^{(-)}_{p}t-\vec{p}.\vec{r})\right)\right]\nonumber\\+\sum_{\vec{k}}\left(\frac{\hbar
c^{2}}{2V\omega^{(+)}_{k}}\right)^{\frac{1}{2}}\left[b(\vec{k})\;exp\left(-\frac{i}{\hbar}(E^{(+)}_{p}t-\vec{p}.\vec{r})\right)
+b^{\dag}(\vec{k})\;exp\left(\frac{i}{\hbar}(E^{(+)}_{p}t-\vec{p}.\vec{r})\right)\right],
\end{eqnarray}
where we shall take the solutions $\Phi(x)$ to lie in a large
cube of side $L$ and volume $V=L^{3}$. The first two terms on the
right-hand side of Eq. (28) for $\beta\rightarrow 0$ will be
converted to the general solution of the ordinary Klein-Gordon
equation [15], while the last two terms on the right-hand side of
Eq. (28) for $\beta\rightarrow 0$ are entirely new and they have
no analog in ordinary Klein-Gordon theory. By putting
$\beta'=2\beta$ into equation (6) and neglecting terms of order
$\beta^{2}$, the isotropic minimal length becomes $(\bigtriangleup
X^{i})_{0}\simeq\hbar\sqrt{5\beta}$. The upper bound for
deformation parameter $\beta$ together with isotropic minimal
length $(\bigtriangleup X^{i})_{0}$ are given in Table 1 for some
neutral mesons according to Eq. (22).
\begin{table}
\begin{small}
\begin{center}
\caption{The upper bound values for deformation parameter $\beta$
and its corresponding isotropic minimal length $(\bigtriangleup
X^{i})_{0}$ (the meson masses are taken from Ref.[16])}
\vspace{0.5 cm}
\begin{tabular}{|c||c|c|c|c|}
\hline &&&& \\ Meson & Quark content & Mass &
$\beta_{_{upper-bound}}\simeq\frac{1}{8m^{2}c^{2}}$ &
$(\bigtriangleup X^{i})_{0}\simeq \hbar
\sqrt{5\beta_{_{upper-bound}}}
$\\ & & $(MeV/c^{2})$& $(10^{36}\frac{s^{2}}{kg^{2}m^{2}})$ & $(10^{-16}m)$ \\
&&&& \\
\hline\hline &&&& \\
$\pi^{0}$ & $(u\bar{u}-d\bar{d})/\sqrt{2}$ & 134.964 & 24.027 & 11.56 \\
$\eta$ & $(u\bar{u}+d\bar{d}-2s\bar{s})/\sqrt{6}$ & 548.8 & 1.453 & 2.84 \\
$\eta^{'}$ & $(u\bar{u}+d\bar{d}+s\bar{s})/\sqrt{3}$ & 957.6 & 0.477 & 1.63 \\
$\eta_{c}$ & $c\bar{c}$ & 2981 & 0.049 & 0.52 \\
&&&& \\
\hline
\end{tabular}
\end{center}
\end{small}
\end{table}

\section{Conclusions}
``In the past few years, a large amount of research work has been
devoted to the study of the minimal length uncertainty relation.
The idea behind this minimal length uncertainty is, if we take
into account the effects of quantum fluctuations of the
gravitational field in order to incorporate gravity into quantum
mechanics, a significant consequence deduced from this
unification is that in quantum gravity, there exists a minimal
observable length of the order of the Planck distance. This
minimal length uncertainty is related to a modification of the
standard Heisenberg algebra by adding small corrections to the
canonical commutation relations [17].'' An immediate consequence
of the minimal length uncertainty relation is a generalization of
momentum operator according to Eq. (11). This generalized form of
momentum operator leads to a fourth-order Klein-Gordon equation.
We have shown that our modified Klein-Gordon equation which
contains fourth-order derivative of the wave function describes
two massive particles, one physical particle with the effective
mass $m_{-}$ and the other a ghost particle with the effective
mass $m_{+}$ according to Eqs. (20) and (21). From Eqs. (20) and
(21), one finds the restriction on the deformation parameter
$\beta:\beta<\frac{1}{8m^{2}c^{2}}$ which leads to the isotropic
minimal length $(\bigtriangleup X^{i})_{0}\simeq
\frac{\sqrt{10}}{4}\frac{\hbar}{mc}$. According to Table 1 the
isotropic minimal length in our analysis lies in the interval
$10^{-17}m<(\bigtriangleup X^{i})_{0}<10^{-15}m$. The above range
for the isotropic minimal length is compatible with the results
of Refs. [8-13]. In [8-12] considering the Lamb shift the authors
estimated $(\bigtriangleup X^{i})_{0}\leq 10^{-16}-10^{-17}m$,
analysis of electron motion in a Penning trap also gives
$(\bigtriangleup X^{i})_{0}\leq 10^{-16}m$ [13]. On the other
hand, consideration of neutron motion in the gravitational field
[18] gives large $(\bigtriangleup X^{i})_{0}\leq
2.4\times10^{-9}m$ due to significant experimental errors. In
Ref. [19], the author studies the standard Casimir effect of two
perfectly conducting plates and obtains $(\bigtriangleup
X^{i})_{0}\leq 15 nm$. In Ref. [20], the effect of minimal length
in the Casimir-Polder interactions between neutral atoms is
studied and the author estimates the minimal length must be in
the range $80nm<(\bigtriangleup X^{i})_{0}<10\mu m$. Note that
the reported values for the isotropic minimal length in Refs.
[18-20] are slightly different from the numerical values of
$(\bigtriangleup X^{i})_{0}$ in our work and
Refs. [8-13].\\

\section*{Acknowledgments}
We would like to thank  Ali Asghar Vafainia for discussions
during early stages of this work.

\section*{Appendix A. Lagrangian Formulation of the Real Klein-Gordon Field in the Presence of a Minimal Length}
\renewcommand{\theequation}{A.\arabic{equation}}
\setcounter{section}{0} \setcounter{equation}{0}

The Klein-Gordon Lagrangian density for a real scalar field is
[16]
\begin{equation}
{\cal
L}(\Phi,\partial_{\mu}\Phi)=\frac{1}{2}g^{\mu\nu}(\partial_{\mu}\Phi)(\partial_{\nu}\Phi)-\frac{1}{2}
(\frac{mc}{\hbar})^{2}\Phi^{2}.
\end{equation}
The Euler-Lagrange equation for the field $\Phi$ is
\begin{equation}
\frac{\partial{\cal
L}}{\partial\Phi}-\partial_{\mu}(\frac{\partial{\cal
L}}{\partial(\partial_{\mu}\Phi)})=0.
\end{equation}
If we substitute the Lagrangian density (A.1) into the
Euler-Lagrange equation (A.2), we will obtain the Klein-Gordon
equation as follows
\begin{equation}
\Box\Phi+(\frac{mc}{\hbar})^{2}\Phi=0.
\end{equation}
So far we have considered Lagrangians which were functions of
field quantities and their first derivatives only. Now we want to
obtain the Lagrangian density for the real Klein-Gordon field in
the presence of a minimal length. For such a purpose, let us
write the Lagrangian density by using the representations (10)
and (11), i.e.,
\begin{equation}
x^{\mu}\rightarrow x^{\mu},
\end{equation}
\begin{equation}
\partial^{\mu}\rightarrow(1+\beta\hbar^{2}\Box)\partial^{\mu}.
\end{equation}
The result reads
\begin{eqnarray}
\nonumber {\cal
L}&=&\frac{1}{2}g^{\mu\nu}\left[(1+\beta\hbar^{2}\Box)\partial_{\mu}\right]\Phi\;\left[(1+\beta\hbar^{2}\Box)\partial_{\nu}\right]
\Phi-\frac{1}{2}(\frac{mc}{\hbar})^{2}\Phi^{2} \\
\nonumber
&=&\frac{1}{2}\left[g^{\mu\nu}(\partial_{\mu}\Phi)(\partial_{\nu}\Phi)-2\beta\hbar^{2}(\Box\Phi)(\Box\Phi)\right]
-\frac{1}{2}(\frac{mc}{\hbar})^{2}\Phi^{2} \\
&&+\partial^{\mu}\left[\beta\hbar^{2}(\partial_{\mu}\Phi)(\Box\Phi)\right]+{\cal
O}(\beta^{2}).
\end{eqnarray}
After neglecting terms of order $\beta^{2}$ and dropping out the
total derivative term
$\partial^{\mu}\left[\beta\hbar^{2}(\partial_{\mu}\Phi)(\Box\Phi)\right]$
, the Lagrangian density (A.6) will be equivalent to the
following Lagrangian density
\begin{equation}
{\cal
L}=\frac{1}{2}\left[g^{\mu\nu}(\partial_{\mu}\Phi)(\partial_{\nu}\Phi)-2\beta\hbar^{2}(\Box\Phi)(\Box\Phi)\right]
-\frac{1}{2}(\frac{mc}{\hbar})^{2}\Phi^{2}.
\end{equation}
The second term on the right-hand side of (A.7) shows the effects
of quantum gravitational corrections. If Lagrangian density ${\cal
L}$ depends on first- and second-order derivatives of the fields,
the Euler-Lagrange equations will take the form [21]
\begin{equation}
\frac{\partial{\cal
L}}{\partial\Phi}-\partial_{\mu}(\frac{\partial{\cal
L}}{\partial(\partial_{\mu}\Phi)})+\partial_{\mu}\partial_{\nu}(\frac{\partial{\cal
L}}{\partial(\partial_{\mu}\partial_{\nu}\Phi)})=0.
\end{equation}
If we substitute the Lagrangian density (A.7) into the
generalized Euler-Lagrange equation (A.8), we will obtain the
Klein-Gordon equation in the presence of a minimal length as
follows
\begin{equation}
\Box\Phi+2\beta
\hbar^{2}\Box\Box\Phi+(\frac{mc}{\hbar})^{2}\Phi=0.
\end{equation}

\section*{Appendix B. Feynman Propagator for the Real Klein-Gordon Field in the Presence of a Minimal Length}
\renewcommand{\theequation}{B.\arabic{equation}}
\setcounter{section}{0} \setcounter{equation}{0}

In order to discuss the Green's functions of the theory we let
the generalized Klein-Gordon field interact with an external
source by adding the term $J(x)$ to the right-hand side in Eq.
(14). The inhomogeneous generalized Klein-Gordon equation becomes
\begin{equation}
\left(\Box+2\beta
\hbar^{2}\Box\Box+(\frac{mc}{\hbar})^{2}\right)\Phi(x)=J(x).
\end{equation}
Then the solution to the wave equation (B.1) is given by
\begin{equation}
\Phi(x)=\Phi_{0}(x)-\int\Delta_{F}^{M}(x-y)J(y)d^{4}y,
\end{equation}
where $\Delta_{F}^{M}(x-y)$ is the modified Feynman propagator
defined by
\begin{equation}
\left(\Box_{x}+2\beta
\hbar^{2}\Box_{x}\Box_{x}+(\frac{mc}{\hbar})^{2}\right)\Delta_{F}^{M}(x-y)=-\delta^{4}(x-y),
\end{equation}
and $\Phi_{0}(x)$ is any function that satisfies the homogeneous
wave equation
\begin{equation}
\left(\Box+2\beta
\hbar^{2}\Box\Box+(\frac{mc}{\hbar})^{2}\right)\Phi_{0}(x)=0.
\end{equation}
On making the Fourier transform,
\begin{equation}
\Delta_{F}^{M}(x-y)=\frac{1}{(2\pi)^{4}}\int d^{4}k \;
D_{F}^{M}(k) \; e^{-ik.(x-y)},
\end{equation}
where $D_{F}^{M}(k)$ is the modified momentum space propagator
and substituting it into (B.3), we get
\begin{eqnarray}
\nonumber
D_{F}^{M}(k)&=&\frac{1}{k^{2}-2\beta\hbar^{2}\left(k^{2}\right)^{2}-\left(\frac{mc}{\hbar}\right)^{2}}
\\
\nonumber &=&\frac{1}{2\beta
c^{2}(m_{+}^{2}-m_{-}^{2})}\left[\frac{1}{k^{2}-\left(\frac{m_{-}c}{\hbar}\right)^{2}}-
\frac{1}{k^{2}-\left(\frac{m_{+}c}{\hbar}\right)^{2}}\right] \\
&=&\frac{1}{\left(1-8\beta
m^{2}c^{2}\right)^{\frac{1}{2}}}\left[\frac{1}{k^{2}-\left(\frac{m_{-}c}{\hbar}\right)^{2}}-
\frac{1}{k^{2}-\left(\frac{m_{+}c}{\hbar}\right)^{2}}\right].
\end{eqnarray}
Equation (B.6) is valid only in the case $m_{+}\neq m_{-}$ or
$\beta\neq\frac{1}{8 m^{2} c^{2}}$. According to Eqs. (22) and
(B.6) we have two particles, one with the effective mass $m_{-}$
and the other a Weyl ghost of effective mass $m_{+}$. The ghost
gives the negative contribution to the energy [22-25] and as a
result, the Hamiltonian is not bounded from below. Therefore, in
order to formulate the quantum field theory we must introduce
indefinite metrics [22-25]. In the limit $\beta\rightarrow 0$,
the modified momentum space propagator $D_{F}^{M}(k)$ in (B.6)
smoothly becomes the conventional momentum space propagator
$D_{F}(k)$ [15], i.e.,
\begin{eqnarray}
\nonumber \lim _{\beta\rightarrow 0} \; D_{F}^{M}(k)&=&D_{F}(k) \\
&=&\frac{1}{k^{2}-\left(\frac{mc}{\hbar}\right)^{2}}\; .
\end{eqnarray}

\end{document}